\documentclass[conference]{IEEEtran}
\IEEEoverridecommandlockouts
\usepackage{cite}
\usepackage{amsmath,amssymb,amsfonts}
\usepackage{algorithm}
\usepackage{algpseudocode}
\usepackage{algorithmicx}
\usepackage{graphicx}
\usepackage{textcomp, soul}
\usepackage{xcolor}
\usepackage{float}
\usepackage[caption=false]{subfig}

\usepackage{ifpdf}
\ifpdf
\usepackage{hyperref}
\fi
\usepackage{url}

\usepackage{booktabs}
\usepackage{array}
\newcolumntype{C}[1]{>{\centering\arraybackslash}m{#1}}
\newcolumntype{L}[1]{>{\raggedright\arraybackslash}m{#1}}

\begin{document}

\title{Pruning Blockchain Protocols for Efficient Access Control in IoT Systems}
	
	\author{\IEEEauthorblockN{Huang, Yongtao}
		\IEEEauthorblockA{\textit{University of Texas at Dallas} \\
			huang.yongtao@utdallas.edu}
		\and
		\IEEEauthorblockN{I-Ling Yen}
		\IEEEauthorblockA{\textit{University of Texas at Dallas} \\
			ilyen@utdallas.edu}
		\and
		\IEEEauthorblockN{Farokh Bastani}
		\IEEEauthorblockA{\textit{University of Texas at Dallas} \\
			bastani@utdallas.edu}
	}
	
\maketitle
\thispagestyle{plain}
\pagestyle{plain}

\begin{abstract}
We consider access control for IoT systems that involves shared accesses to the IoT devices as well as their data. Since IoT devices are dispersed all over the edge of the Internet, traditional centralized access control has problems. Blockchain based decentralized access control is thus the new solution trend. However, existing blockchain based access control methods do not focus on performance issues and may incur a high communication overhead. 

In this paper, we develop a Pruned Blockchain based Access Control (PBAC) protocol to cutdown the unnecessary message rounds and achieve high efficiency in access validations and policy management. The protocol includes a shortcut and a Role and Device Hierarchy-Based Access Control (R\&D-BAC) approaches for different environment settings. To realize the PBAC protocols, it is necessary to carefully engineer the system architecture, which is also discussed in the paper.  Experiments demonstrate the efficacy of the PBAC protocol, specifically, the shortcut mechanism reduces access time by approximately 43\%, and R\&D-BAC outperforms traditional blockchain based RBAC by more than two folds. 

\end{abstract}


\section{Introduction}
Internet of Things (IoT) is growing in volume as well as importance. An IoT system connects a diverse range of devices, from wearable, home, industrial, surveillance, agricultural, and smart city sensors to mobile devices, autonomous vehicles, drones, etc., and is becoming indispensable in enhancing the convenience and efficiency of our daily operations. This rapid expansion of IoT yields great social benefits, but at the same time, introduces security challenges. 


Current IoT systems in practice typically use basic encryption-based handshakes or Single Sign-On (SSO) gateways for access control  \cite{hardt_oauth_2012}. 
While handshakes are simple, they lack the complexity needed for sharing resources effectively. Conversely, SSO relies on a stable internet connection, often leaving users without access during outages, a notable issue in smart homes where unreachable remote server-managed authentication can block accesses that is completely local. Moreover, some manufacturers control device management through proprietary social networks or apps, requiring account registration and ``family'' joins, which centralizes control and raises security and privacy concerns. Advanced users may circumvent these issues with solutions like dynamic DNS, reverse proxies, and SSL certificates, but these can be complex and insecure for the average user.

Several cloud vendors, such as AWS IoT Core \cite{aws_secure_2024}, Microsoft Azure IoT Hub \cite{microsoft_azure_iot_2024} and Oracle IoT \cite{oracle_iot_2024}, also provide IoT system platforms to facilitate the computation on the edge for IoT devices. However, most of these platforms are extensions of their existing cloud platforms. The access control models in some of these cloud-extended systems are more sophisticated, supporting policy-based access control. However, the system structure for managing access control in these systems is still centralized. Thus, users still need to go to the central certificate authority (or named differently but with the same functionality) to get access validation and authorization.

Due to the pervasive nature of the IoT systems, centralized access control has the inherent communication overheads as well as the local access disruption problem during Internet disconnection period. In the latter case, users cannot access IoT devices on the same local network. Thus, it is essential to consider decentralized access control. 
Decentralized access control has also been considered in the literature.
In \cite{sicari_policy_2017}, access control for smart health systems is considered. It addresses the cross domain issue and proposes to use the registration process to ensure that users and devices have global unique IDs. However, how the domain specific policies are defined such that these universal IDs can be used for access validation against policies from different domains have not been given. 
\cite{liu_access_2017} considers role based cross domain access control issues in manufacturing IoT (MioT) systems. The focus of the paper is to identify the optimal authorization route with the least spread permissions. A PGAO algorithm is developed for finding the optimal authorization route. The proposed approach can provide an efficient authorization decision process in multiple collaborative domains in MIoT. This work does not consider how to perform cross domain role mappings, but assumes that such mapping rules are already defined in advance. 
Also, the authorization path finding is a centralized solution, requiring all the users, roles, permissions, and mapping policies of the involved domains being maintained at a central site in order to run the PGAO algorithm. Also, the mechanism cannot be used in dynamic situations since role mappings are defined in advance. 
\cite{alkhresheh_daciot_2020} addresses the adaptivity issue in highly dynamic IoT environment and proposes a dynamic access control framework, DACIoT. The framework extends the XACML model to include three core components: Automatic Policy Specification (APS), Continuous Policy Enforcement (CPE), and Adaptive Policy Adjustment (APA). APS breaks down access control rules into fundamental elements and their contextual conditions. So, the access control policy can be updated in a finer grain. CPE monitors changes in the operation context and general context during access sessions. APA then uses anomaly detection techniques to identify misbehavior and adjust access control policies on the fly according to session contexts. 

The earlier decentralized access control works have a strong sense of large domains and the domain security units and cross domain security management are assumed to be reliable and trustworthy. With large-scale IoT systems consisting of numerous large and small domains as well as individualized IoT device owners, access control  based on predefined cross-domain policies becomes infeasible and more dynamic schemes and less trustworthy assumptions should be considered. With these concerns, blockchain based access control has been proposed.
Most of the works in blockchain based access control embed access control in established blockchain platforms, including public blockchain platforms, such as Bitcoin and Ethereum, and consortium blockchain platforms, such as Hyperledger Fabric \cite{androulaki_hyperledger_2018} and FISCO-BCOS \cite{li_fisco-bcos_2023}. Different platforms result in different architectures, performance, and embedding mechanisms. 

Earlier blockchain based access control works, such as \cite{ouaddah_fairaccess_2016} and \cite{chen_blockchain_2017} are Bitcoin based.
In general, there are several shortcomings in Bitcoin based access control. First, the transaction protocol is predefined, therefore, special encoding scheme is required to embed customized information into the blockchain transactions. Also, the access control solutions requires digital currency for recording transactions. The transaction history is stored on all bitcoin peers and, hence, wasting space for non-participators. The slow block update time of Bitcoin, approximately 10 minutes, and its CPU-intensive Proof-of-Work (PoW) consensus method render Bitcoin unsuitable for time-sensitive work.

On Bitcoin and Consortium blockchains, the policy embedding mechanism is transactional where creation and updating of access control policies are issued as transactions to the blockchain. On the other hand, Ethereum based access control embed policies as smart contracts, making use of the contract languages offered by Ethereum. 
There are many Ethereum based access control frameworks being proposed and most of them focus on architecture designs. \cite{alphand_iotchain_2018} relies on the IoT devices themselves for managing accesses. This approach may have problems because some IoT devices do not have sufficient computation and storage capabilities for handling blockchain activities. On the other hand, \cite{novo_blockchain_2018}, \cite{almadhoun_user_2018}, and \cite{zhang_smart_2019} attempt to balance scalability and the potentially limited capabilities of IoT devices and use some mediator nodes to manage the interactions with the blockchain for the IoT devices they are in charge of. But the mediator nodes may fail and cannot ensure continuous operations. 
Overall, none of the approaches specify how to encode access control policies into smart contracts. Such encoding can be difficult for expressing some XACML rules, such as timing related ones, though the contract language is Tuning complete. 
A more serious problem with Etheruem based solutions is still the performance concerns. It greatly improves over Bitcoin and has around 12 seconds latency. Also, the requirement of digital currency still exists. Moreover, the execution time of smart contracts is untrackable because which miners would accept the request and whether the miners will win the competition are unknown.

Consortium blockchain involves a preselected set of validators instead of public miners for managing the blockchain. Also, it does not rely on Proof-of-Work or Proof-of-Stake to achieve consensus among all peers. Instead, Practical Byzantine Fault Tolerance (PBFT) between dedicated validators is sufficient and can yield a much shorter update time. Moreover, most consortium blockchains do not require any cryptocurrency, avoiding unnecessary involvement of a large number of miners for maintaining the chain. Thus, it incurs least overhead.
\cite{ding_novel_2019} is an ABAC based IoT access control framework built on top of Hyperledger Fabric. \cite{tan_blockchain-empowered_2021} is deployed on FISCO-BCOS \cite{li_fisco-bcos_2023}. The transactions recorded on blockchain in a form analogous to Access Control List (ACL). 
Though the performance of consortium blockchain based access control is relatively reasonable, much better than that of public blockchains, there is still rooms for improvements. 

We develop a pruned blockchain based access control (PBAC) protocol for IoT systems, which prunes the unnecessary interactions between the IoT devices, their AC hubs, and the blockchain. We also engineered the underlying blockchain support based on a bespoke consortium blockchain specifically optimized for access control. The PBAC protocol includes bootstraping, access request handling, and access policy validation algorithms. As with many other blockchain based access control solutions, 
PBAC also provides a robust crash recovery mechanism that is specific to our protocol, making it versatile for use in large-scale IoT systems with IoT resource sharing.

The PBAC pruning mechanisms include a shortcut approach that can be used for IoT systems with permanent and long term users and a role and device hierarchy-based access control (R\&D-BAC) for organizations with many IoT devices organized in a hierarchy.

For access request handling, we require that both the device domain and the blockchain service to validate the access rights of the request against the policies. With a little additional overhead due to duplicated processing, we enable a significant reduction in the number of message rounds via a shortcut mechanism, which allows full access validation being done in parallel with local access authorization. This can benefit accesses by permanent users, such as the owner, and by recurrent accessors, which occur frequently in real-world access patterns.

To further enhance the performance in blockchain based access control, we consider dealing with accesses to IoT devices organized in a hierarchy. Resource hierarchy is widely adopted by many major IoT providers. Since users frequently have the tendency to access resources that are correlated, e.g., in a subtree of a device hierarchy, we design a novel R\&D-BAC access control model and design the associated access validation mechanism to greatly reduce the number of message rounds required for blockchain related validation.

Experimental results show that shortcut accesses, compared with normal full path accesses, shorten the latency by over 40\%. The R\&D-BAC approach further improves the performance of blockchain based RBAC by more than two folds.

\section{System Overview}\label{pbac:sec:arc}

\subsection{System Entities}
\begin{figure}[htbp]	\centering\noindent\includegraphics[width=0.48\textwidth]{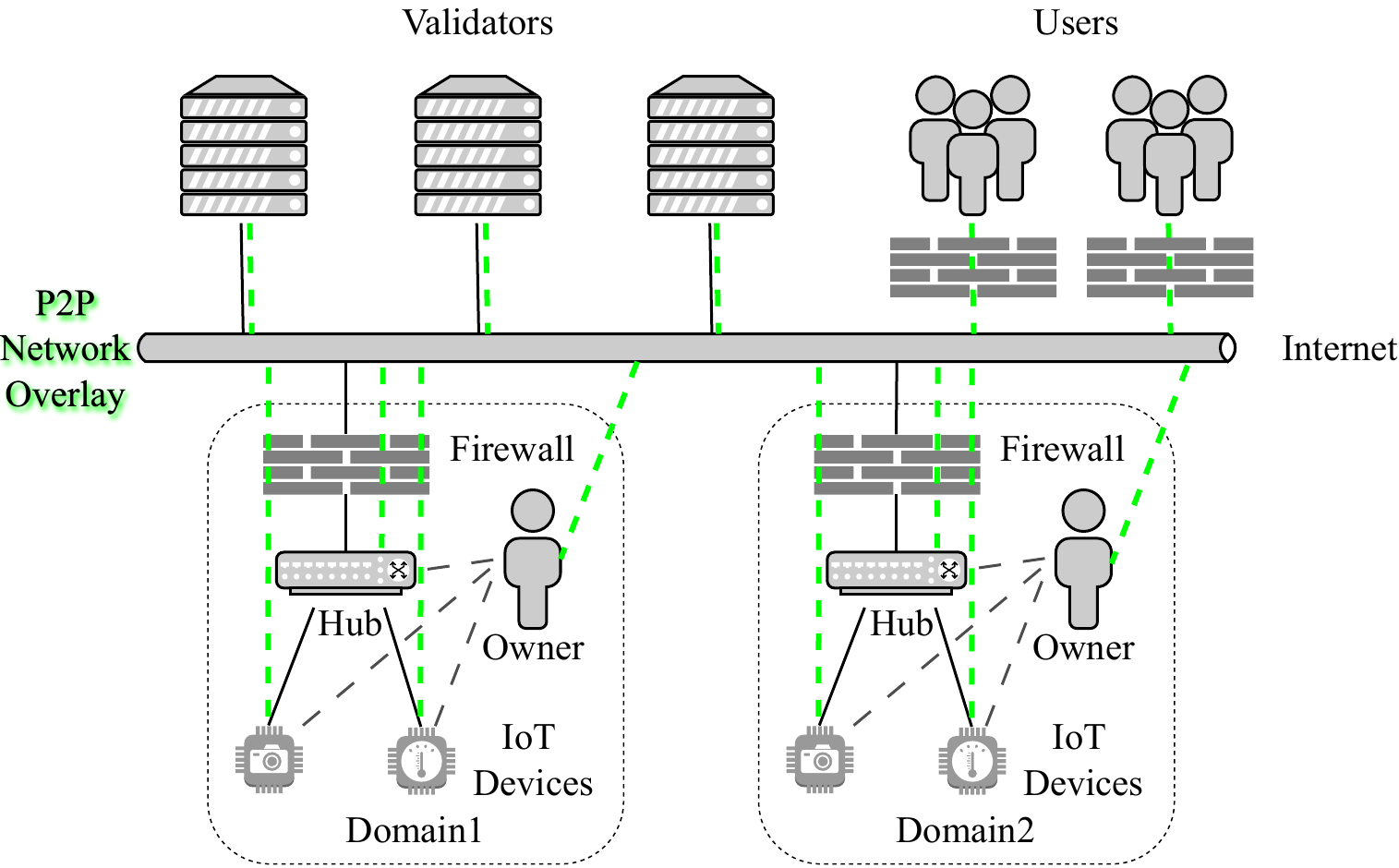}
	\caption{System Architecture}
	\label{pbac:fig_overall}
\end{figure}

The important entities involved in our blockchain based access control protocols are depicted in Figure \ref{pbac:fig_overall} and each entity is further defined and elaborated subsequently. 

\textbf{IoT Devices} offer a wide spectrum of IoT services. As how they are placed in the current Internet, we consider them hidden in the subnets behind firewalls.
Services provided by the IoT devices are defined via manufacturer-provided templates, which can be used for semantic based service discovery.
Each IoT device can be the basic element for access control and policies can be assignable on a per-service basis.

\textbf{Users} can access the IoT devices in the network according to the access control polices for the devices. A user can represent a diverse range of entities, including an independent individual, an individual in a domain, a software program or a service, or even some IoT entity or its service. 

\textbf{Owners} Device owner can be an individual, an organization or another device in hierarchical setup. The entity who owns the device has full access rights to the device. Ownership is immutable once device is registered.
%


\textbf{Domains} A domain is a higher-level entity in the system that consists of a group of IoT devices. IoT devices and their owners shall be in the same domain. A single IoT device with sufficient capabilities can choose to form its own domain. Many IoT devices are limited in memory and storage, rendering them incapable of performing complex operations. To enable access control on constrained devices not belonging to specific administrative units, they can be grouped together to form a domain managed by more powerful AC hubs. In many cases, domains are established for certain administrative units, such as homes, factories, companies, organizations, etc. In these cases, the domain may host a generic policy for the devices or device groups in the domain instead of defining policies for individual IoT devices. Frequently, the IoT devices in the domain form a resource hierarchy and access control policies can be defined based on the hierarchy (will be discussed later). 

\textbf{Validator Cluster} 
A cluster of validators are used to realize the blockchain based decentralized access control. To avoid performance overhead, we consider a consortium of blockchain validators, consisting of dedicated servers provided by trustworthy IoT vendors. These validators are pre-certified by the consortium's security standards to ensure their trustworthiness (Proof-of-Authority). Though rarely, individual validators may fail or be compromised, a consensus protocol is used to ensure the trustworthiness of the cluster.

Validators maintain the blockchain's complete state and are responsible for adding new blocks, using the Ditto\cite{gelashvili_jolteon_2022} consensus protocol—a 2-chain HotStuff\cite{yin_hotstuff_2019-1} variant—for network consensus. The data structure, based on the JellyFish Merkle tree\cite{gao_jellyfish_2021} from the Diem project\cite{amsden_libra_2020}, is optimized for storage efficiency and I/O overhead.

\textbf{Access Control Hubs}
Each domain has an AC hub to manage access control and blockchain related tasks for the IoT devices in the domain. For operational efficacy, the AC hubs maintain the local domain state, including up-to-date access control policies for all connected devices, and the recent states necessary for their functionality.

Note that some domain formations are not based on specific organizations, thus, the domain hubs may not always be trusted. Also, for organizational domains, we consider that their domain AC hubs may be compromised and become untrusted.


\subsection{PBAC Bootstrapping}\label{pbac:ssec:boot}
Access control policies for each domain in the system are codified into the blockchain via domain registration. An authorized entity of a domain must first execute a registration transaction to join the system and associating its devices and owners with the domain. (Note that a domain could be a single device with sufficient computation and storage capabilities.) The access control policies of the domain are established subsequently via blockchain transactions by the authorized entities, such as the corresponding owners or security officers of the domain. 
Domain and device registration related transactions are specified as follows:

\begin{itemize}
	\item\textbf{Domain Registration} involves recording the creator's ID $id_{issuer}$, domain ID $uid_{domain}$, owner's ID $uid_{owner}$, and a policy descriptor $pd$ submitted in a transaction $(id_{issuer}, uid_{domain}, uid_{owner}, pd)$.  The policy descriptor either points to a predefined common access control model, such as models we will discuss in the later sub-sections, or a user defined policy through smart contract.
	\item\textbf{Device Registration} should be applied equally to AC hubs and IoT devices (an IoT device may serve as its own AC hub). The transaction is represented by $(id_{issuer}, uid_{domain}, did, uid_{owner}, \{sv, \cdots\})$, where $uid_{domain}$ and $uid_{owner}$ are as defined earlier and $\{sv, \cdots\}$ is the list of service names.
	\item\textbf{Device Revocation} with parameters $(id_{issuer}, did)$ nullifies a registered device $did$.
\end{itemize}

Once registered, the device record in blockchain is immutable. Changes in ownership require releasing by the current owner and reregistration by the new owner. 

\section{Access Validation in PBAC}\label{pbac:ssec:val}
In existing works that embed access control policies in the block-chain \cite{zhang_smart_2019} \cite{ding_novel_2019}, access validation is always performed at the validator cluster or miners. We call this the \textbf{full path} validation. This remote validation can introduce significant delays. To shorten the latency for access validation, we consider a basic pruning technique in PBAC, the \textbf{shortcut} protocol, for accesses to devices by trusted users, such as the owners and pre-registered long-term users.
In the shortcut protocol, validation is executed at the AC hub of a domain first and the full validation at the validator is done in parallel in the background. The benefit of the shortcut protocol is not only the greatly reduced access latency, but also allowing continued accesses in local areas when the internet connection is disrupted. This can benefit many local IoT systems, such as home IoT systems and IoT systems in factories, farms, hospitals, small companies, etc. The blockchain embedded access control would have been undesirable for local IoT systems or even infeasible for critical local systems while our shortcut protocol, with simple pruning, makes it highly applicable.

\textbf{ShadowDP Data Structure} To facilitate the realization of the shortcut protocol, the access control policies for individual devices and domains should not only be logged in the blockchain, but also maintained at the AC hub. AC policy changes will not only be reflected at the validator cluster, but also at the domain AC hub. For organizational domains, this policy maintenance is natural and changes generally happen at the domain AC hubs anyway. For domains with low power devices of independent administrative units, the up-to-date policies are also maintained by the corresponding AC hubs. We consider the hub copy of the policies as the \textbf{shadowDP}.
Also, in existing blockchain based access control, the access control policies are only be logged as transactions in the blockchain, including the original establishment and the subsequent modifications of the policies. Thus, the validators need to rebuilt the policies from these logs to get the view of the policies, which can be highly inefficient during access validation. To address this problem, we extend the \textbf{shadowDP} to the validator side. Each validator maintains indices to access control (AC) policy entries in the blockchain and keeps another shadow copy of the most recent policies in memory to ensure efficient access validations.

\begin{figure}[h]
	\centering\noindent%
	\subfloat[Full Path]{\includegraphics[width=0.23\textwidth]{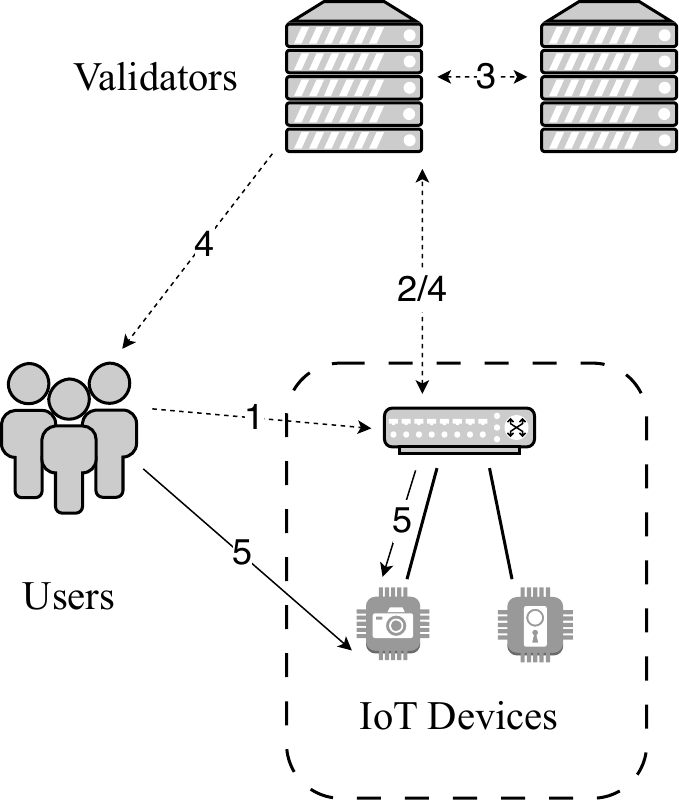}}\hfil%
	\subfloat[Internet Shortcut]{\includegraphics[width=0.23\textwidth]{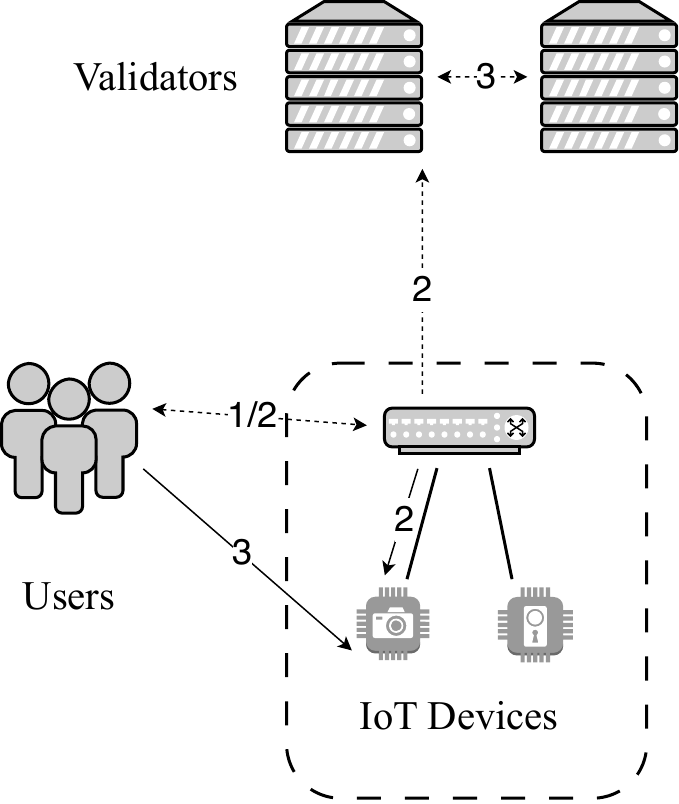}}\hfil%
	\subfloat[Local Shortcut\label{pbac:fig_local}]{\includegraphics[width=0.23\textwidth]{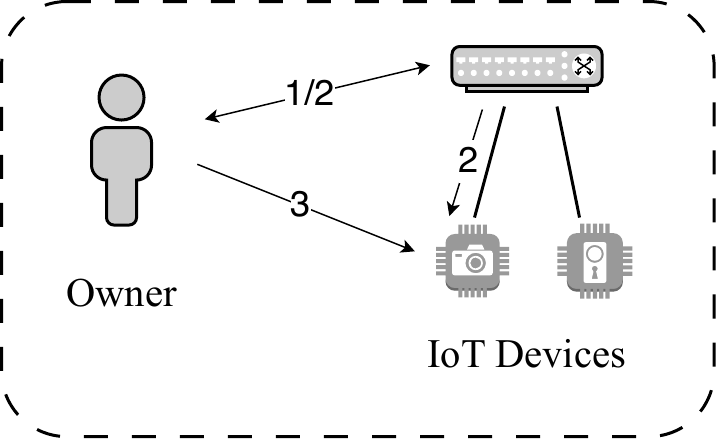}}
	\caption{Full Path and Shortcut Access}\label{pbac:fig_vali}
\end{figure}

The shortcut versus the full path access validation procedure is depicted in Figure \ref{pbac:fig_vali}.

\textbf{Full Path (normal) access validation} 
If a user with identity $uid$ trying to access an IoT device with identity $did$ and its corresponding service $sv$ (optional) with permission type $pt$ for the first time, then it issues the access request $req=(uid, did, pt, sv)$ to the AC hub of $did$ (which could be $did$ itself). The AC hub, with identity $hid$, upon receiving $req$, validates $req$ against its access control policy $ACP$, and after successful validation, generates a permission $per=(uid, did, pt, sv, et, ul)$ and sign it into a token $t=s_{hid}(uid, did, pt, sv, et, ul)$, where $et$ is the optional expiration time, $ul$ is the optional usage limitations, and $s_{hid}$ is the signature for the token $t$. Token $t$ is then forwarded to the Validator Cluster for endorsement. Validators, after successfully double validating the access rights for permission $per$ against $ACP$ in its shadowDP, perform Byzantine agreement (BA) to reach consensus on $per$, and add $t$ to the blockchain with majority's signature $s_{v_i\cdots, hid}(uid, did, pt, sv, et', ul')$ and return the endorsement. After getting the endorsement from a majority of Validators, the AC hub creates a session key $k(uid, did, pt, sv, et', ul')$ and sends it to the device $did$ which enables $uid$'s access via $t$. Also, token $t$ is returned to $uid$ to allow its access.

If $uid$ has accessed $did$ before and still has the key-value pair $(req,t)$ in its cache (where $t$ is the corresponding token for $req$), $uid$ can simply use $t$ for direct re-entry  without requiring further validation, unless the token expires or the session key is lost due to insufficient memory, power outages, or reboots on $did$. The use of paired token (for user) and session key (for device) expedites repeated accesses. 

\textbf{Shortcut Access Validation} In the shortcut protocol, a domain AC hub maintains a list $T_u$, which records such users. The AC hub, upon receiving accesses from users in $T_u$, sends out the token to the accessor without waiting for the endorsement by Validators to complete (which is done in parallel instead).  In the highly unlikely occasions of failed endorsement, remedy actions such as reconducting the endorsement or revoking the session key as well as other mitigation steps can be taken to protect the system. Also, the use of ECDSA key pairs supports long token expiration time and facilitates shortcut and other designs to achieve high performance gains.  

The PBAC access validation algorithms by the requester, the AC hub, and the Validators are illustrated in Figure \ref{pbac:fig_vali} and detailed in Algorithms \ref{pbac:alg:req}, \ref{pbac:alg:vali}, and \ref{pbac:alg:endo} (modifications captured in them will be discussed later).

\begin{algorithm}\caption{User $uid$ requests to connects to device $did$}\label{pbac:alg:req}
	\begin{algorithmic}[0]
		\Procedure{request}{$req=(uid, did, pt, sv)$}
		\If{search in $uid$'s cache with key $req$ reutrns $t$}
		\State reuse $t$ to request connection to $did$
		\If{connection established}
		\State proceed to access $did$
		\State exit procedure
		\EndIf
		\EndIf\Comment{No token or session key expires}
		\State send $req$ to AC hub $hid$
		\State wait for token $t=s_{hid}(uid, did, pt, sv, et, ul)$
		\If{$t$ has been received}
		\State cache key-value pair $(req,t)$
		\State use $t$ to request connection to $did$
		\If{connection established}
		\State proceed to access $did$
		\Else
		\State temporary failure, retry later
		\EndIf
		\EndIf
		\EndProcedure
	\end{algorithmic}
\end{algorithm}
%

\begin{algorithm}\caption{AC hub $hid$ validates request $req$}\label{pbac:alg:vali}
	\begin{algorithmic}[0]
		\Procedure{validate}{$req=(uid, did, pt, sv)$ }
		\If{$req$ satisfies $ACP$ in shadowDP} 
		\State \Comment{$ACP$ is the access control policy}
		\State find the corresponding $(et_p, ul_p)$ in $ACP$
		\State $per\leftarrow(uid, did, pt, sv, et_p, ul_p)$
		\State $t \leftarrow s_{hid}(per)$
		\If{$uid\in T_u$}\Comment{Shortcut}
		\State send session key $k(per)$ to $did$
		\State send token $t$ to $uid$
		\State multi-cast $s_{v_i, hid}(per)$ to Validator Cluster
		\Else\Comment{Full path}
		\State multi-cast $s_{v_i, hid}(per)$ to Validator Cluster
		\State wait for endorsed token $s_{v_i\cdots, hid}(per))$ 
		\State {\hspace{50px} from majority of Validators}
		\State send session key $k(per)$ to $did$
		\State send token $t$ to $uid$
		\EndIf
		\Else
		\State send $reject$ to $uid$
		\EndIf
		\EndProcedure
	\end{algorithmic}
\end{algorithm}

\begin{algorithm}\caption{Validator endorses the token}\label{pbac:alg:endo}
	\begin{algorithmic}[0]
		\Procedure{endorse}{$s_{hid}(per)$}
		\If{$req$ satisfies $ACP$ in shadowDP} 
		\State $v_i$ multi-cast $s_{v_i, hid}(per)$ to other validators
		\If{BFT consensus is archieved}
		\State $v_i$ sends $s_{v_i\cdots, hid}(per)$ to AC hub $hid$ 
		\State Commit to blockchain
		\EndIf
		\EndIf
		\EndProcedure
	\end{algorithmic}
\end{algorithm}


\textbf{Light workload for IoT devices and Continued accesses upon Internet disconnection} The access control validation algorithm is designed to minimize the workload on the IoT devices, The devices are not involved in the validation procedure and only need to maintain the session keys and accordingly accept or reject the incoming accesses.

In scenarios where internet connectivity is lost as shown in Figure \ref{pbac:fig_local}, our system supports intranet accesses, allowing token validation and device access within a local network. This capability ensures continuous operations, with synchronization to the blockchain resumed upon connectivity restoration.

\section{PBAC for Various Access Control Models}
To illustrate the compatibility of PBAC, we integrate several established models, including Discretionary Access Control (DAC), Attribute-Based Access Control (ABAC), and Role-Based Access Control (RBAC), with PBAC in Subsections \ref{pbac:ssec:dac}, \ref{pbac:ssec:rbac} and \ref{pbac:ssec:abac}. For each model, the permission transaction structure and the specific revocation and recovery mechanisms are discussed.

\subsection{Discretionary Access Control (DAC) in PBAC}\label{pbac:ssec:dac}
In DAC, permissions are structured in Access Control List (ACL) at the device hub. Each ACL entry records granted permission tuples that mirror blockchain transactions. ACL and its index are stored in ShadowDP. Invalid entries due to expired $et$ or now violated access limitation $ul$ are purged by monitoring timers.

Blockchain captures permission transactions instead of the full ACL. Each transaction record captures the ID of the issuer, the permission tuple, as well as $et$ and $ul$. Different types of transactions and their fields are formally defined as follows.
\begin{itemize}
	\item\textbf{Permission Granted} $(id_{issuer}, uid, did, pt, sv, et, ul)$
	\item\textbf{Permission Revoked} $(id_{issuer}, uid, did, pt, sv, et, ul)$
\end{itemize}

Revoking access involves dispatching a revocation session key to the device. Following acknowledgment, the corresponding permission entry is expunged from the ACL.

The sequential recording in the blockchain inherently supports the reconstruction of ACLs on AC hubs by replaying pertinent permission transactions since domain registration. Although a centralizing ACL could pose risks in an enterprise setting, synchronizing ACLs across multiple AC hubs within the same domain enhances fault tolerance. This flexible architecture permits the addition or removal of AC hubs as needed, without impacting the overall system integrity and functionality.


\subsection{Role-Based Access Control (RBAC) in PBAC}\label{pbac:ssec:rbac}
In a RBAC domain, permissions are aligned with user roles rather than individual identities, represented by the tuple $(rid, did, pt, sv)$. Here $rid$ is a universally unique identifier assigned to each role, distinct from libp2p network peer identities. The uniqueness of $rid$ is verified through validator authentication upon role creation. Roles may also be associated with user-friendly names ($rn$), which are encrypted by default to protect privacy across domains, given that cross-domain access utilizes $rid$.
For simplicity and without loss of generality, we omit the optional  fields $et$ and $ul$.

Transactions specific to RBAC for role creation, modification, and deletion are itemized as follows.
\begin{itemize}
	\item\textbf{New Role} $(id_{issuer}, id, rid, rn)$ where $id$ is domain ID.
	\item\textbf{Delete Role} $(id_{issuer}, rid)$
	\item\textbf{Assign Role User} $(id_{issuer}, rid, uid)$
	\item\textbf{Remove Role User} $(id_{issuer}, rid, uid)$
	\item\textbf{Assign Role Permission} $(id_{issuer}, rid, did, pt, sv)$
	\item\textbf{Revoke Role Permission} $(id_{issuer}, rid, did, pt, sv)$
\end{itemize}

Each AC hub maintains a bidirectional mapping between users and their respective roles ($map_{ur}={uid: [rid, \cdots]}$ and $map_{ru}={rid: [uid, \cdots]}$) in ShadowDP. When a user requests access, the AC hub retrieves the user's role $rid$ and verifies if the requested permission $(rid, did, pt, sv)$ is authorized. 
Adding a user to a role involves updating both $map_{ur}$ and $map_{ru}$ to reflect the new association. Conversely, removing a user from a role necessitates the issuance of a revocation session key to the relevant device(s) before adjusting the mappings, mirroring the revocation process in DAC.

The elimination of a role necessitates a comprehensive re-evaluation of permissions for all users associated with that role, considering that an individual's access rights might still be valid through other roles they hold. Below is an algorithm detailing this re-evaluation process.
\begin{algorithm}\caption{Re-evaluation of user permissions}\label{pbac:alg:eval}
	\begin{algorithmic}[0]
		\Function{re-evaluate}{$rid$ }
		\State $original\leftarrow$ all permission set of $rid$ in ACL
		\ForAll{$u\in map_{ru}[rid]$}
		\State $new\_set\leftarrow \emptyset$
		\ForAll{$r\in map_{ur}[u]$}
		\State add permission set of $r$ to $new\_set$
		\EndFor
		\State $revoked\_set\leftarrow (original-new\_set)$
		\ForAll{$d\in revoked\_set$}
		\State send revocation session key
		\EndFor
		\EndFor
		\EndFunction
	\end{algorithmic}
\end{algorithm}

Similar to the DAC model, an AC hub administering RBAC policies can reconstruct its state by replaying recorded permission transactions. This capability ensures that the system can recover from disruptions and maintain consistent access control policies, even after hub failures or network disruptions.

\subsection{Attribute-Based Access Control (ABAC) in PBAC}\label{pbac:ssec:abac}
ABAC extends the functionality of DAC by incorporating a versatile attribute list and context-based access policies. Similar to RBAC, attributes in ABAC are assigned a unique identifier and can have a user-friendly name ($aname$), which is encrypted by default to ensure privacy. However, ABAC differentiates itself by allowing attributes to be associated with both devices and users, with permissions defined by the tuple $(aid_{user}, aid_{device}, pt, sv)$, where $aid$ represents the attribute identifier for either the user or device. Different types of transactions for ABAC are enlisted in the following.
\begin{itemize}
	\item\textbf{New Attribute} $(id_{issuer}, id, aid, aname)$. $id$ is domain ID.
	\item\textbf{Delete Attribute} $(id_{issuer}, aid)$
	\item\textbf{Assign Attribute Device} $(id_{issuer}, aid, did, sv)$
	\item\textbf{Remove Attribute Device} $(id_{issuer}, aid, did, sv)$
	\item\textbf{Assign Attribute User} $(id_{issuer}, aid, uid)$
	\item\textbf{Remove Attribute User} $(id_{issuer}, aid, uid)$
	\item\textbf{Assign Attribute Permission}\\ $(id_{issuer}, aid_{user}, aid_{device}, pt)$
	\item\textbf{Revoke Attribute Permission}\\  $(id_{issuer}, aid_{user}, aid_{device}, pt)$
\end{itemize}

The access validation and assignment procedures and the reconstruction of access control policies in an ABAC system mirror those established in RBAC and is not discussed further.

\section{Role and Device Hierarchy-Based Access Control (R\&D-BAC)}\label{pbac:sec:rd}
In enterprise settings, where devices and users proliferate, they are often organized hierarchically to support more systematic access control policy definitions. Though there are well defined access control theories and models for role hierarchy, the access control models for device hierarchy is lagging. 

Within a device hierarchy (DH), an interior node generally represents a device group owned by a subdomain in the organization. In a role hierarchy, permissions granted to an ancestor node are extended to its descendants. Similarly, we consider that in a DH, granting permissions for accessing a device group implicitly implies that the same permissions are granted for accessing all the devices in the subtree (subdomain).

The device hierarchy based access control model can greatly simplify access right assignments. Instead of assigning access rights to a large number of individual devices, domain/subdomain administrators can assign desired access permissions to a device group at any level in the DH. Upper level subdomains in the DH offer more limited access rights to more limited accessor groups, while additional permissions can be assigned to the lower level subdomains closer to the actual devices.

\begin{figure}[h]
	\centering\noindent%
	\subfloat[\textit{control}\label{pbac:acn}]{\includegraphics[width=0.2\textwidth]{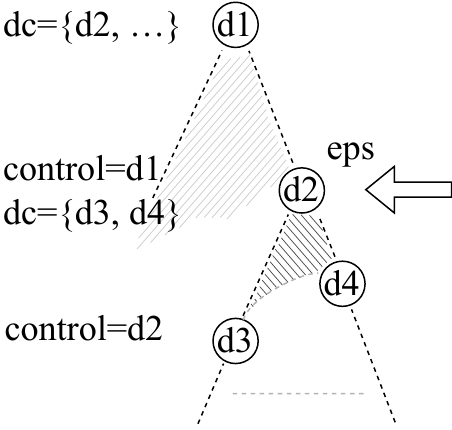}}\hfil%
	\subfloat[\textit{non-control}\label{pbac:nonacn}]{\includegraphics[width=0.23\textwidth]{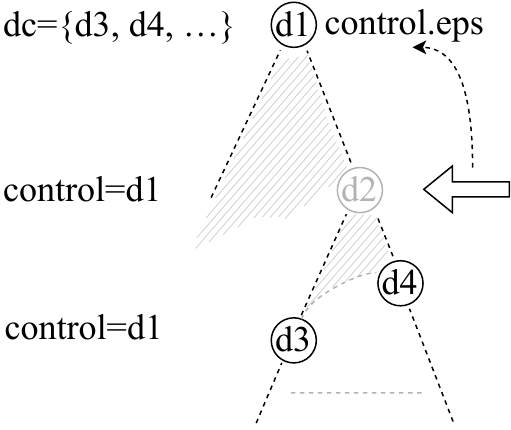}}
	\caption{Visit \textit{control} and \textit{non-control} Device}\label{pbac:fig:control}
\end{figure}

\textbf{Control Devices and Control Pointers} Upon assigning a permission tuple $(rid, did, pt, sv)$ to a device $did$ in the DH, the designated device becomes a control device. Devices without direct permissions look up to this control device for access validation. Correspondingly, each node maintains a control pointer ($did.control$) to its closest control device in its ancestry. Additionally, a control device keeps track of a list of pointers $did.dc$, pointing to the control devices within its closest descendants along different paths. 

As illustrated in Figure \ref{pbac:acn}, device $d_2$ functions as a control device. Its ancestor control node pointer $d_2.control$ points to $d_1$. Also, $d_3.control$ and $d_4.control$ point up to $d_2$. On the other hand, $d_1.dc$ contains $d_2$ and $d_2.dc$ contains $d_3$ and $d_4$. 

Control devices cache both the permissions directly assigned to them ($did.ps$) and a cumulative permission set inherited from their ancestor control devices ($did.eps$), where $did.eps$ includes $did.ps$.
When a user seeks permission to access a control device (e.g., $d_2$), the device reviews its effective permission set $eps$ and, upon finding the request permissible, accept the access. Conversely, if the accessed device is not a control device, it refers to its control device pointer (e.g., $d_2.control=d_1$). The request is then relayed to its control device to determine the request's validity and communicates the outcome accordingly. This procedure is outlined in Algorithm \ref{pbac:alg:rd-val}.

\begin{algorithm}\caption{Device validates permission}\label{pbac:alg:rd-val}
	\begin{algorithmic}[0]
		\Procedure{validate}{$req=(uid, did, pt, sv)$ }
		\If{ControlValidate(req)}
		\State $t\leftarrow s_{did}(per)\leftarrow$ sign $per$\Comment{Access granted}
		\State store session key $k(per)$
		\State multi-cast $t$ to validators or forward $t$ to AC hub
		\State response $t$ to $uid$
		\Else
		\State response $reject$
		\EndIf
		\EndProcedure
		
		\Function{ControlValidate}{$req=(rid, did, pt, sv)$ }
		\If{$did.ps\neq \emptyset$}\Comment{control device}
		\State\Return check $req\in did.eps$
		\Else\Comment{non control device}
		\State\Return $did.control.validate(per)$\Comment{check its control device for result}
		\EndIf
		\EndFunction
	\end{algorithmic}
\end{algorithm}

\begin{figure}[h]
	\centering\noindent%
	\subfloat[Insertion\label{pbac:ins_ctl}]{\includegraphics[width=0.24\textwidth]{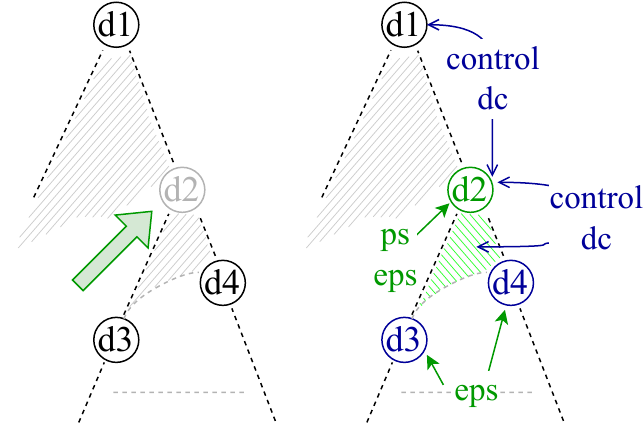}}\hfil%
	\subfloat[Deletion\label{pbac:rem_ctl}]{\includegraphics[width=0.24\textwidth]{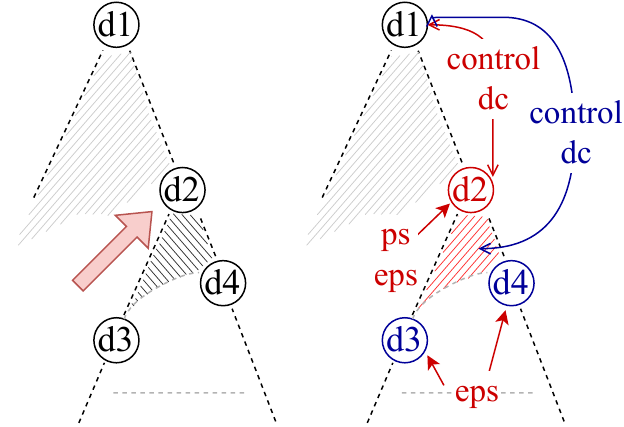}}
	\caption{Insert and Remove Control Device}\label{pbac:fig:insdel}
\end{figure}

Assigning permission to a non-control device transforms it into a control device, whereas revoking the last permission from a control device reverts it to a non-control status. This transition is visually represented in Figures \ref{pbac:fig:insdel} and outlined in Algorithm \ref{pbac:alg:rd-control}.

\begin{algorithm}\caption{Transformation of control state}\label{pbac:alg:rd-control}
	\begin{algorithmic}[0]
		\Function{InsertControl}{$did$}
		\State add $did$ to $did.control.dc$
		\ForAll{$d\in descendants\ of\ did$}\Comment{Search down device tree}
		\State $d.control\leftarrow did$
		\If{$d.ps\neq \emptyset$}\Comment{Control device}
		\State add $d$ to $did.dc$
		\State remove $d$ from $did.control.dc$
		\State stop searching down
		\EndIf
		\EndFor
		\EndFunction
		
		\Function{RemoveControl}{$did$}
		\State remove $did$ from $did.controld.dc$
		\State empty $did.eps$
		\ForAll{$d\in descendants\ of\ did$}
		\State $d.control\leftarrow did.control$
		\If{$d.ps\neq \emptyset$}\Comment{Control device}
		\State add $d$ to $did.control.dc$
		\State remove $d$ from $did.dc$
		\State stop searching down
		\EndIf
		\EndFor
		\EndFunction
	\end{algorithmic}
\end{algorithm}

Permission assignment may involve propagating the permission down along the control device tree as depicted in Algorithm \ref{pbac:alg:rd-assi}.
\begin{algorithm}\caption{Permission assignment}\label{pbac:alg:rd-assi}
	\begin{algorithmic}[0]
		\Function{assign}{$per=(rid, did, pt, sv)$ }
		\If{$did.ps=\emptyset$}
		\State $InsertControl(did)$\Comment{To control device}
		\EndIf
		
		\If{$per\not\in did.ps$}
		\State add $per$ to $did.ps$
		\State $PropagateEps(did)$
		\EndIf
		\EndFunction
		
		\Function{revoke}{$per=(rid, did, pt, sv)$ }		
		\If{$per\in did.ps$}
		\State remove $per$ from $did.ps$
		\State $PropagateEps(did)$
		\EndIf
		\If{$did.ps=\emptyset$}
		\State $RemoveControl(did)$\Comment{To non-control device}
		\EndIf
		\EndFunction
		
		\Function{PropagateEps}{$did$}\Comment{Traverse down control devices tree}
		\State $new\_eps\leftarrow did.ps\cup did.control.eps$
		\If{$did.eps\neq new\_eps$}
		\State $did.eps\leftarrow new\_eps$
		\ForAll{$d \in did.dc$}
		\State $PropagateEps(d)$
		\EndFor
		\EndIf
		\EndFunction
	\end{algorithmic}
\end{algorithm}

For example, in the configuration depicted Figure \ref{pbac:ins_ctl}, $d_1$, $d_3$ and $d_4$ serve as control devices, with $d_2$ initially a non-control device. Upon the assignment of permission to $d_2$, which changes it to control device, adjustments are made to the control device tree to reflect this change. Specifically, $d_2$ is added to $d_1.dc$ while $d_3$ and $d_4$ are removed from $d_1.dc$ and added to $d_2.dc$. The control pointers for all devices situated between $d_2$, $d_3$ and $d_4$ inclusive that previously pointing to $d_1$ are redirected to $d_2$.

This conversion of $d_2$ into a control device, necessitates the revision of its $ps$ and $eps$ with the latter being disseminated down to $d_3$ and $d_4$. Consequently, the $eps$ of $d_3$ and $d_4$ are updated to mirror these changes. The process of reverting a control device back to a non-control status operates in a manner analogous to its initial conversion.

Upon the registration of a new device at the leaf level, its $control$ pointer is set to its immediate parent control device. If the parent is a non-control device, the $control$ pointer is then set to mirror the $control$ pointer of the parent. When revoking a control device, it is removed from it $control.dc$.

The transaction protocol, policy recovery and management of roles within this context follows the same principles as established in RBAC. Control devices hold only segments of the role-user ($map_{ur}$) and user-role ($map_{ru}$) mappings relevant to their permissions $eps$. Updates to these mappings are efficiently disseminated across the network through a control device tree

\section{Experimental Study}\label{pbac:sec:perf}

\subsection{Environment Setup}
We utilized commodity hardware to simulate a real-world IoT system scenario with the following entities.
\begin{itemize}
	\item\textbf{Validators Platform} In our experiment, we set up 4 validators, each is an Intel NUC Skull Canyon with a quad-core i7-6770HQ processor, 64GB RAM, and a 2TB M.2 SSD.
	
	\item\textbf{AC Hub} utilizes a virtual server platform, which is a Synology DS1522+ NAS, powered by a dual-core Ryzen R1600 processor with 8GB RAM, of which 1GB is allocated for the virtual machine.
	
	\item\textbf{IoT Device} is simulated by a Raspberry Pi 3 B+, featuring a quad-core BCM2837B0 processor and 1GB RAM.
	
	\item\textbf{Client} A Macbook Pro is used for client operations.
\end{itemize}

This array of devices is networked together using both wired and wireless connections to a central switch, with internet access provided via a router and cable modem. Additionally, a cellphone with AT\&T hotspot serves as a secondary internet access point, enabling us to simulate internet-to-internet communication.

\subsection{Network Setup and Latency Testing}
\textbf{Network Initialization} For the initial network connection, bootstrapping employs techniques like Distributed Hash Table (DHT), Zero Configuration Networking (zeroconf), or rendezvous points from a static list or previous peerstore. Although bootstrap time varies, it's a one-time background process not factored into latency assessments.

\textbf{Firewall Navigation} P2P connections navigate firewalls using manual or uPnP port forwarding, hole punching, or fallback methods when hole punching fails. While port forwarding offers immediate connection, hole punching—requiring two communication stages after a temporary relayed connection\cite{seemann_decentralized_2022}—presents variability in connection times. Libp2p development tests reveal average connection times of 0.89s and up to 7.78s for successful P2P connections\cite{ipfs_libp2p_2022}. Due to its significant variability, our experiments primarily utilize uPnP for firewall traversal.

\textbf{Latency Measurements} Libp2p developer tests show pub-sub communication latency typically stays below 150ms, peaking at 165ms ($P_{99}$) in a 1000-node network using gossipsub-v1.1\cite{vyzovitis_gossipsub-v11_2020}. Local network round trip times are generally less than 1ms.
The raw pub-sub testing involves ``cold start'', that each time a P2P communication is established, peer cache is purged and new peer ID is generated. When considering real-world scenarios, there is no need to cold start every communication. Specifically in our experiments, only the client process is restarted every time, not the AC hubs or the validators. Thus, we redo the latency evaluation for the pub-sub communication considering our cases.

\begin{figure}[ht]
	\centering\noindent\includegraphics[width=0.47\textwidth]{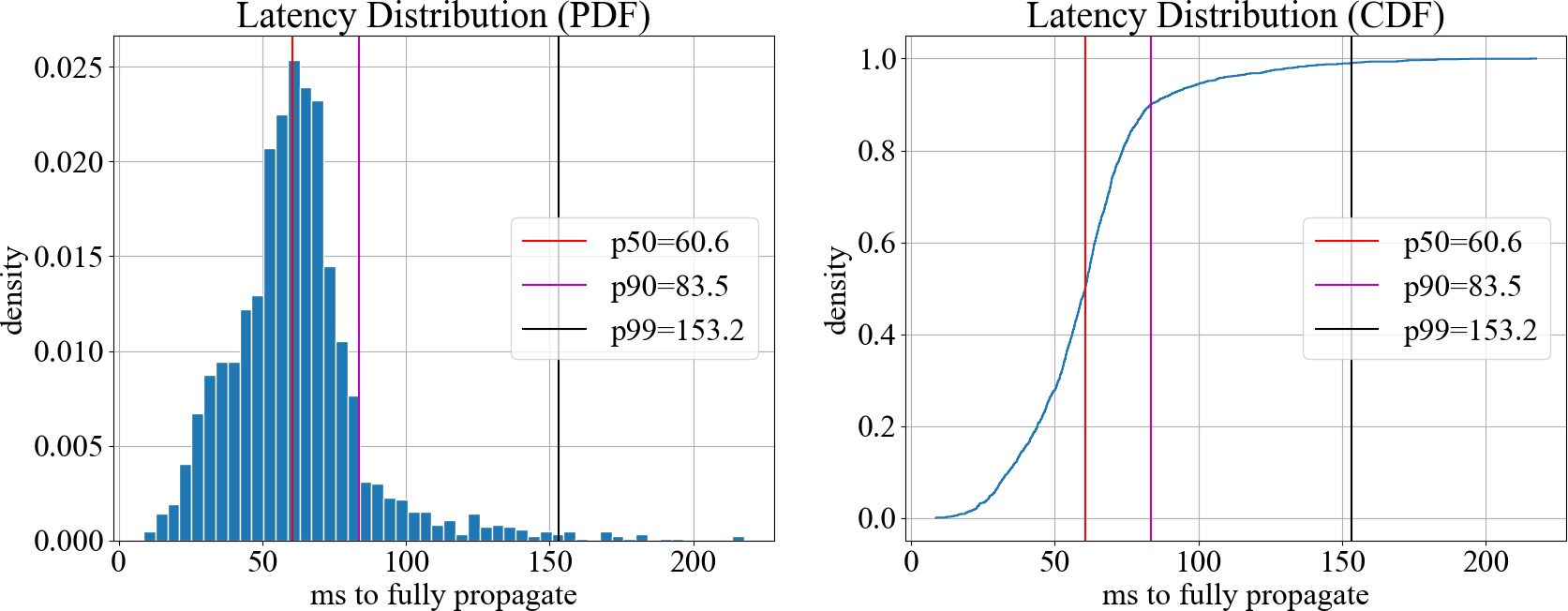}
	\caption{Pub-sub Latency Distribution (Cold Start)}
	\label{pbac:fig_pubsub}
\end{figure}

\begin{figure}[ht]
	\centering\noindent\includegraphics[width=0.47\textwidth]{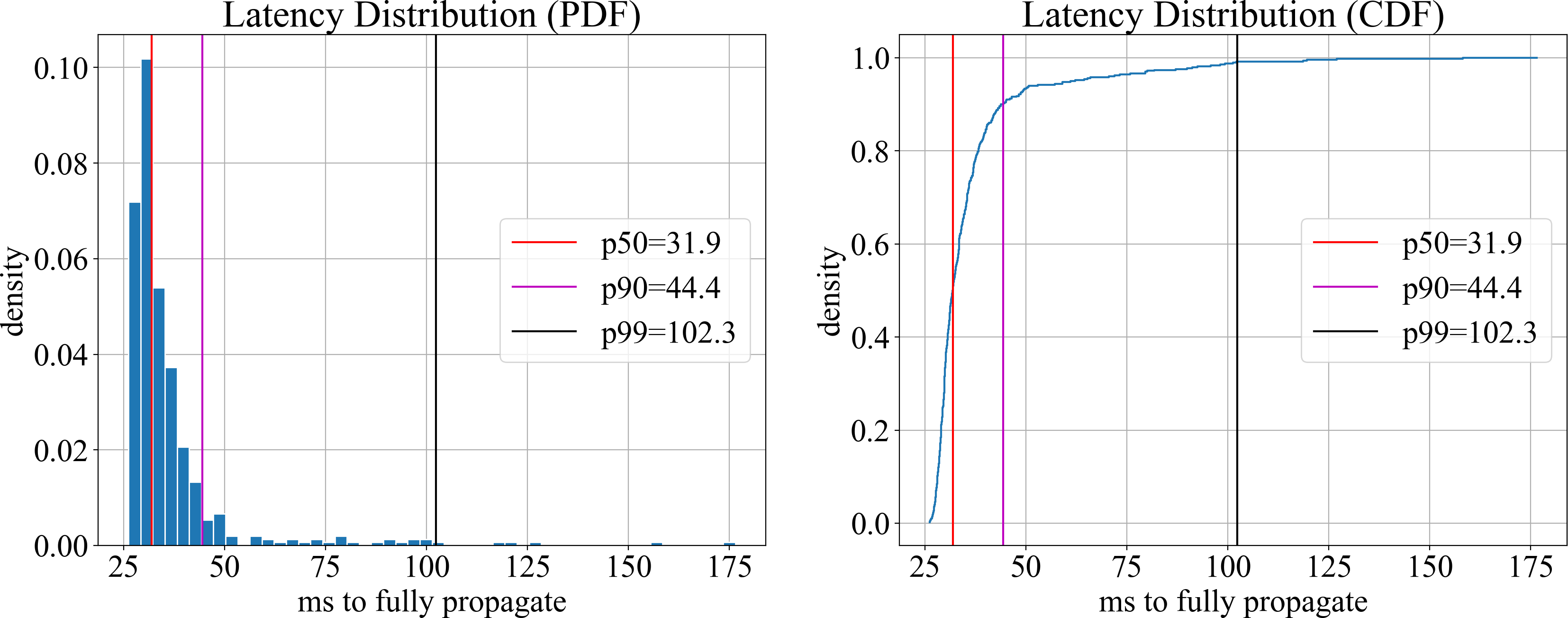}
	\caption{Pub-sub Latency Distribution (Convergence)}
	\label{pbac:fig_p2p}
\end{figure}

We test the latency of internet-to-internet pub-sub communication for 2,000 messages, comparing the case of continuous operations without cold start and the case of cold star for testing peers, i.e., after each process run, we terminated it, and then restarted it, simulating independent connections. The latency distributions for with and without cold start are depicted in Figures \ref{pbac:fig_pubsub} and \ref{pbac:fig_p2p}, respectively.

as shown in Figure \ref{pbac:fig_p2p}, continuous operations without cold start led to a convergence of delivery time, indicating an underlying network optimization that enhances communication efficiency.

\subsection{ Access Validation Latency: Full Path versus Shortcut}
\begin{figure}[ht]
	\centering\noindent\includegraphics[width=0.47\textwidth]{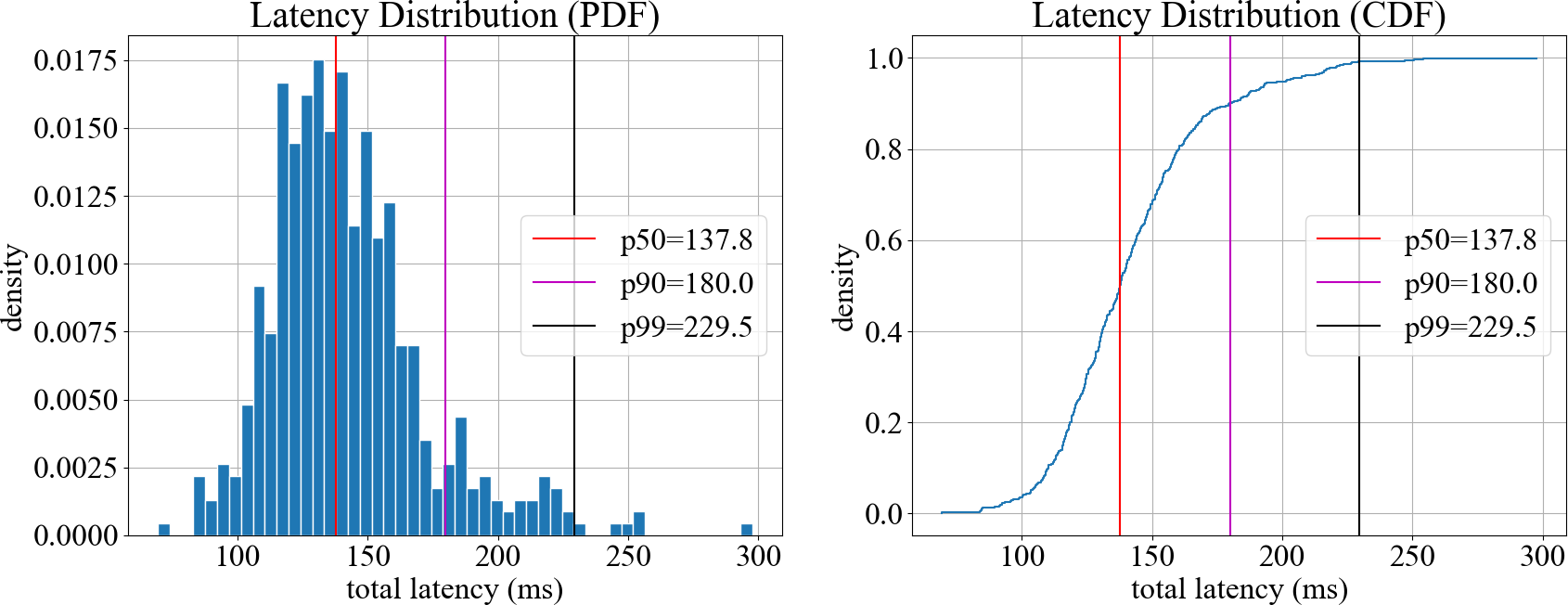}
	\caption{Full Path Access Latency Distribution}
	\label{pbac:fig_full}
\end{figure}

\begin{figure}[ht]
	\centering\noindent\includegraphics[width=0.47\textwidth]{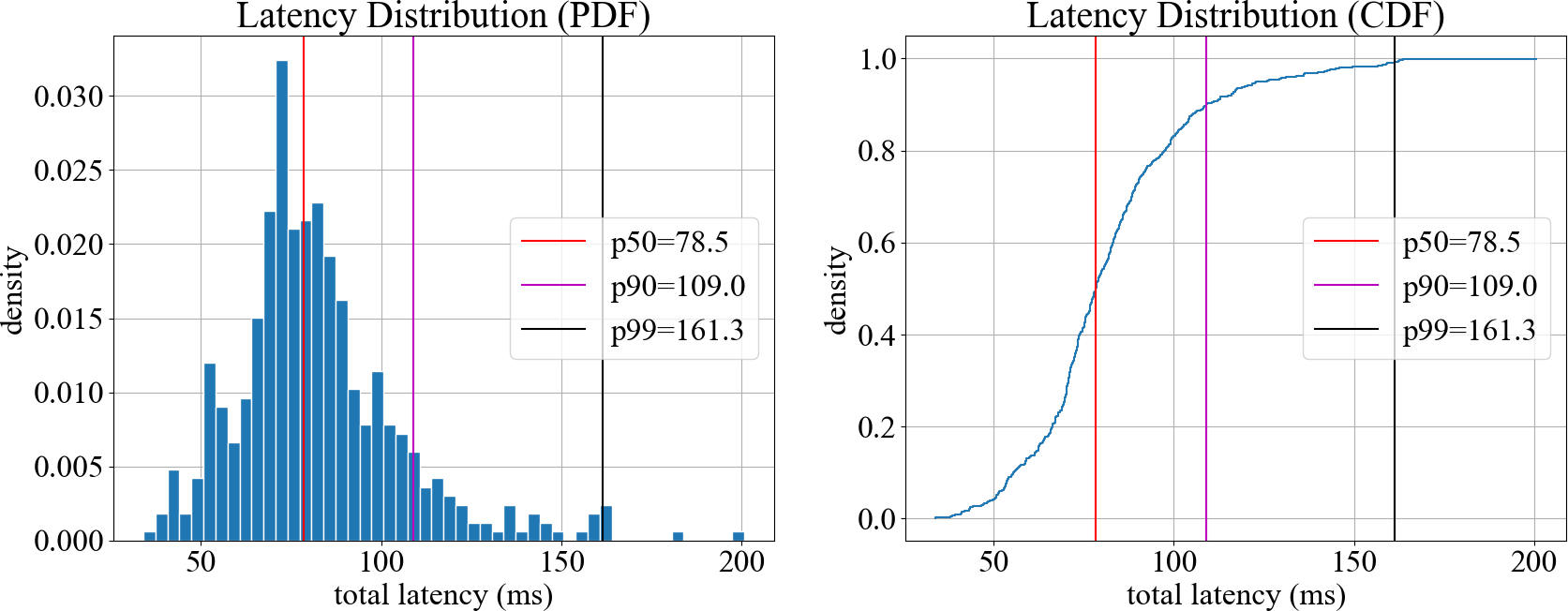}
	\caption{Shortcut Access from Internet Latency Distribution}
	\label{pbac:fig_short}
\end{figure}

\begin{figure}[ht]
	\centering\noindent\includegraphics[width=0.47\textwidth]{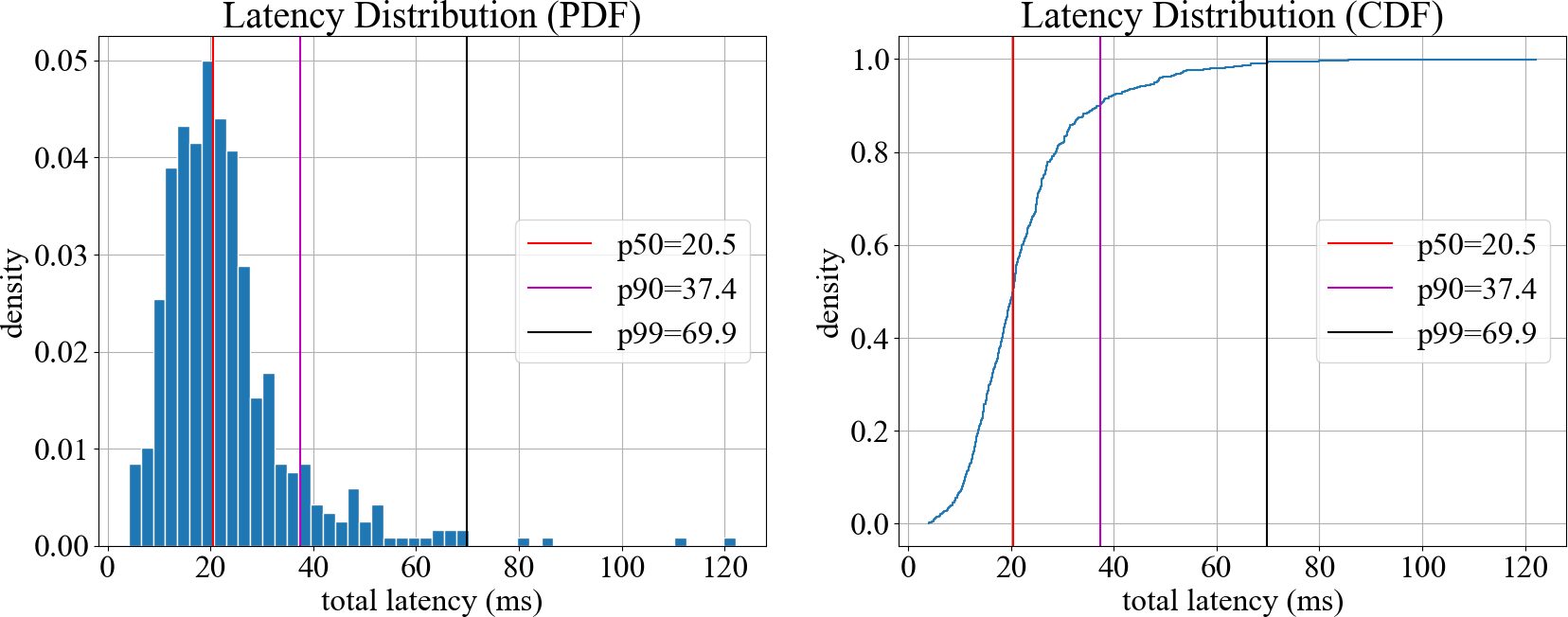}
	\caption{Shortcut Access from Intranet Latency Distribution}
	\label{pbac:fig_short2}
\end{figure}

As discussed earlier, we do not cold start the AC hub and validators, only client process is restarted every time. Figures \ref{pbac:fig_full}, \ref{pbac:fig_short} and \ref{pbac:fig_short2} represent the latency distributions of 500 times of access validations conducted via full path, shortcut from internet, and shortcut from intranet, respectively. 


The results from the figures show that in the median case ($P_{50}$), utilizing shortcut access over the internet yields a 43\% reduction in access time, while intranet shortcuts offer saving of approximately 73.9\% compared to internet shortcuts. In the worst-case scenario ($P_{99}$), internet shortcut access provides a savings of about 29.7\%, and leveraging intranet shortcuts further saves approximately 56.7\%. 

\subsection{Access Validation Latency of R\&D-BAC}
To evaluate the performance of R\&D-BAC, we compare it with the baseline case, i.e., RBAC without device hierarchy. In the baseline case, for each device access of $did$, if a token for accessing $did$ is cached by the accessor $uid$, $uid$ establishes the session with $did$ directly; otherwise, it executes the normal access procedure, directing the request to the AC hub of $did$. 

The experiments are carried out through a simulator, which simulates local network and processing time in R\&D-BAC domain. The DH in an R\&D-BAC domain is simulated via a random tree generator based on specified parameters, including tree \textbf{depth ($H$)} and \textbf{degree ($D$)}, which are the number of layers and the average number of child nodes in the DH, correspondingly. The number of child nodes of each node is generated in a Poisson distribution with $\lambda=D$. In our experiment, we choose $H=10$ and $D=20$.

We randomly generate permission assignment requests to simulate the control nodes in the device hierarchy. The designated nodes for the permission assignment requests are selected based on a normal distribution, with mean ($\mu=0.6$) and standard deviation ($\sigma=0.1$). This setup skews permission assignments towards the middle to lower depths of the device hierarchy (0 for root and 1 for leaf).

\begin{figure}[ht]
	\centering\noindent\includegraphics[width=0.47\textwidth]{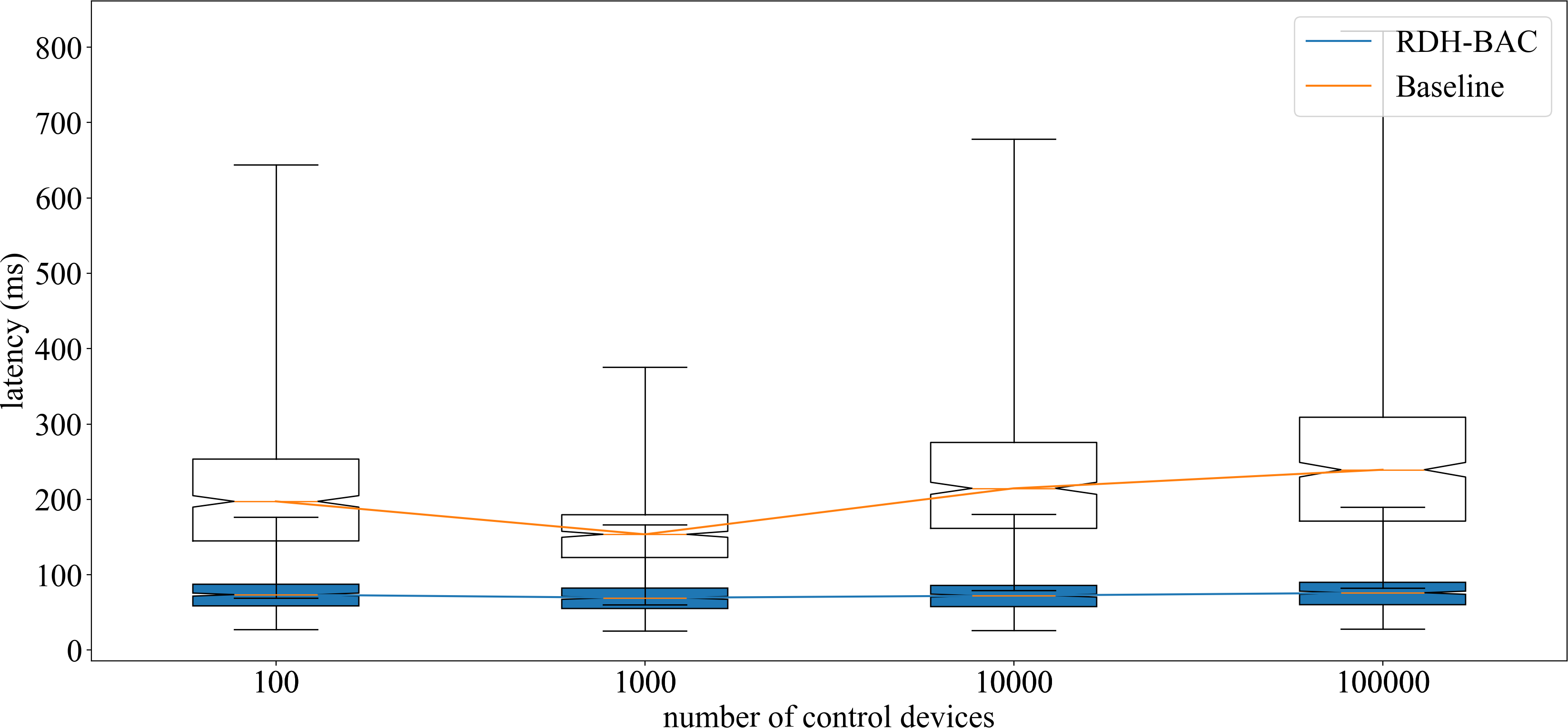}
	\caption{Assignment Latency}
	\label{pbac:fig_acc}
\end{figure}

Access validation latency was evaluated based on the configuration with 1,000,000 devices, 100 roles, and 100,000 users. The performance impact was measured by consider different numbers of control devices, ranging from empty to fully populated: 100, 1,000, 10,000 and 100,000. This allowed us to observe and quantify the effect of scaling the number of control devices on the R\&D-BAC efficiency and responsiveness. The experimental results are presented in Figure \ref{pbac:fig_acc}.

As demonstrated in Figure \ref{pbac:fig_acc}, the R\&D-BAC scheme consistently outperforms traditional RBAC access control methods. On average, R\&D-BAC achieves a latency time saving of 63.8\% ($P_{50}$), and in the 99th percentile of cases ($P_{99}$), it delivers saving of 71.7\%.

\section{Conclusion}
In this paper, we consider efficiency in blockchain based access control and introduce the design of our PBAC protocols. PBAC focuses on pruning the full scale blockchain based access control validation procedure and introduces a shortcut protocol as well as the novel R\&D-BAC approach. The shortcut protocol can greatly reduce access latency without sacrificing security. R\&D-BAC leverages the device hierarchy concept and offers a new access control model. Accordingly, a strategy built on top of R\&D-BAC model for reducing the access validation latency due to blockchain has been developed. 
A thorough performance study for the PBAC protocols has been conducted and results show that it achieves significant performance improvements compared to the traditional blockchain based access control schemes.

We plan to conduct further research in blockchain based access control for IoT systems. First, we will continue to explore performance improvement methods in blockchain protocols for access control. For example, We plan to expand our R\&D-BAC scheme to efficient protocols for the attribute based access control model. Also, we plan to develop new token based authorization schemes that can be used for accessing multiple devices.
Second, we plan to explore the expressiveness in embedding policies in blockchains. We will investigate the mechanisms for encoding XACML policies in blockchain transactions. We will also develop schemes for automated translation of XACML policies to smart contract languages. For both cases, we will investigate the potential limitations in such embedding.
Third, we will explore blockchain based cross domain access control. In large-scale IoT systems consisting of a very large number of domains with different security administrations, defining federated cross-domain policies or pair-wise cross-domain policies in advance is infeasible. We will explore models for dynamic cross-domain policy mappings. 

\bibliographystyle{IEEEtran}
\bibliography{IEEEabrv,Thesis}

\end{document}